\documentclass{aip-cp}

\usepackage[numbers]{natbib}
\usepackage{rotating}
\usepackage{graphicx}

\begin{document}

\title{Gamma Rays From Blazars}

\author[aff1]{Fabrizio Tavecchio\corref{cor1}}

\affil[aff1]{INAF-Osservatorio Astronomico di Brera, Via Bianchi 46, 23807 Merate, (LC), Italy}
\corresp[cor1]{fabrizio.tavecchio@brera.inaf.it}

\maketitle

\begin{abstract}
Blazars are high-energy engines providing us natural laboratories to study particle acceleration, relativistic plasma processes, magnetic field dynamics, black hole physics. Key informations are provided by observations at high-energy (in particular by {\it Fermi}/LAT) and very-high energy (by Cherenkov telescopes). I give a short account of the current status of the field, with particular emphasis on the theoretical challenges connected to the observed ultra-fast variability events and to the emission of flat spectrum radio quasars in the very high energy band. 
\end{abstract}

\section{INTRODUCTION}

Blazars \cite{UrryPadovani1995} represent a quite small but remarkably interesting fraction of the entire population of active galactic nuclei (AGN). Their defining phenomenology includes the presence of a compact unresolved radio core, with flat or even inverted spectrum, extreme (both in timescale and in amplitude) variability at all frequencies (but generally being more extreme at the highest frequencies), high degree of optical and radio polarization. The most distinctive feature, however, is the intense emission in the $\gamma$-ray band, often dominating the bolometric radiative power output. 

Indeed, blazars are the most numerous extragalactic $\gamma$-ray sources, both at GeV and TeV energies. 
The 4-years catalogue of {\it Fermi}-LAT (3FGL, \cite{3FGL}) reports more than 1600 blazars, to be compared with about 30 non-blazar extragalactic sources. Similar is the situation at higher energies. Cherenkov telescopes detected 62 blazars and only 6 non blazar sources (4 radiogalaxies and 2 starburst galaxies). Blazars are further divided in two subgroups, namely BL Lacertae objects, characterized by extremely weak or even absent emission lines in their optical spectra, and  Flat Spectrum Radio Quasars (FSRQ), showing broad emission lines typical of quasars. FSRQs are generally more powerful than BL Lacs and their radiative output tend to be more dominated by the high-energy emission. BL Lacs, on the other hand, are characterized, on average, by larger energies of the emitted $\gamma$-ray photons and, in fact, are the large majority of the blazars detected at VHE.

The peculiarities of blazars are explained assuming that these sources are AGN hosting a jet of plasma 
expelled at relativistic speeds (bulk Lorentz factors $\Gamma\approx 10-20$), whose axis points almost toward the Earth \cite{BlandfordRees1978}. In this geometry, the luminosity of the non-thermal continuum produced within the jet appears amplified by 3-4 orders of magnitude because of  relativistic beaming effects and it easily outshines any other isotropic emission component associated to the active nucleus (accretion disk, emission lines, dust) or to the host galaxy. In the so-called unified model for radio-loud AGN, BL Lacs and FSRQ are the aligned counterpart of FRI (low power) and FRII (high power) radiogalaxies, respectively \cite{UrryPadovani1995}.

The remarkably smooth SED of blazars, extending over the entire electromagnetic spectrum, from  the radio band to $\gamma$-ray energies, is characterized by a typical ``double humped'' shape. The low energy component -- peaking, depending on the source, between the IR and the UV-soft X-ray band -- is understood as beamed synchrotron radiation of relativistic electrons (or, more generally, $e^{\pm}$ pairs), while for the origin of the second component --  reaching its maximum in the $\gamma$-ray band -- there is no complete consensus.
In leptonic models (see e.g., \cite{Maraschi1992}), the emission is explained as inverse Compton (IC) radiation from the same leptons producing the low-energy component. In hadronic scenarios \cite{Boettcher2013}, instead, $\gamma$ rays are thought to originate  from high-energy hadrons (protons) loosing energy through synchrotron emission (if the magnetic field is large enough, \cite{Aharonian2000}) or photo-meson reactions \cite{Mannheim1993}. In the latter case neutrino emission from the decaying charged pions is also foreseen.

Despite  huge efforts, several crucial aspects of the blazar phenomenology remains unclear 
and poorly understood. Even the physical process(es) responsible for the acceleration of the emitting relativistic particles is (are) not very clear. While until few years ago it was widely accepted that the main acceleration mechanism is the Fermi I-like process acting at shock fronts in the flow (diffusive shock acceleration), this view is now strongly debated. Another point subject to lively discussions is the location of the regions from which a large fraction of the observed radiation originates. This problem is particularly acute for FSRQ, in which the radiative environment surrounding the jet (influencing the IC process and causing the absorption of the $\gamma$-ray photons) varies with the distance from the central supermassive black hole. All these discussions have been recently triggered or revitalized by important observational results obtained in the past years by space ({\it AGILE}, {\it Fermi}) and ground-based (Cherenkov arrays) $\gamma$-ray detectors, in particular the evidence for ultra-fast ($\approx$ minutes) variability events and the detection of FSRQ at high ($E>10$ GeV) and very-high ($E>100$ GeV) energies.
A quite interesting discussion concerns also the nature of the so-called {\it extreme} BL Lacs, showing extremely hard TeV continua and limited (if any) variability at high energy (e.g. \cite{Bonnoli2015} and references therein).  It is also important to remind that, besides the astrophysical issues, the intense high-energy photon beams of blazars are ideal probes for the cosmological fields permeating the Universe (the extragalactic background light and the extragalactic magnetic field) or even to search for new particles (axion-like particles) or look for violation of the Lorentz invariance at high energies. For a discussion of these topics see De Angelis (these proceedings).

In the following, after a sketch of the general framework, I will review in particular the observational status concerning the ultra-rapid variability and the VHE emission of FSRQ and the impact on our understanding of the functioning of blazars. 

\subsection{THE GENERAL FRAMEWORK AND ITS EPICYCLES}

Blazar jets are hosted by an active nucleus comprising a central supermassive black hole ($M_{\rm BH}=10^8-10^9$ $M_{\odot}$)  accreting matter from the surroundings. The phenomenological division between FSRQ and BL Lac objects can be interpreted as reflecting a more fundamental difference in the nature of the accretion flow in the two kind of sources, ultimately regulated by the accretion rate of the infalling material, e.g. \cite{CavaliereD'Elia2002,GMT09}. In FSRQ, which show bright thermal features (optical lines) and, in some cases, a bump at  optical-UV frequencies (thought to mark the direct emission from the hot accreting gas), the accretion likely occurs through a radiatively efficient (optically thick) accretion disk. The luminous UV continuum emitted by the disk is responsible for the photoionization of the gas confined in ``clouds" rapidly orbiting the black hole and occupying the so-called broad line region (BLR). Various methods (in particular the reverberation mapping technique) allows us to locate the clouds at $\approx$0.1 parsec (the ``radius" of the BLR), which displays a clear dependence on the luminosity of the disk \cite{Bentz2006}. Farther out (1-10 pc), dust grains -- likely organized in the geometrical shape of a torus -- intercepts a fraction $\xi\approx 0.5$ of the disk continuum, reprocessing it as thermal IR emission (with temperature close to that corresponding to the sublimation of dust, $T\approx 10^3$ K). On the other hand, the lack of strong thermal components in BL Lac optical spectra is generally  interpreted as an evidence that the accretion flow present in these sources is radiatively inefficient, as expected for accretion rates much smaller than that of quasars, when the accretion flow assumes the structure of an ADAF/ADIOS \cite{CavaliereD'Elia2002}.

This scheme could allows one to explain the difference between the GeV $\gamma$-ray spectra of BL Lacs (generally displaying hard spectra) and those of FSRQ (characterized by soft -- photon index larger than 2 -- spectra), as the effect of the different radiative losses characterizing the high-energy electrons in the two kind of sources \cite{GMT09}.
More generally, the interplay between radiative losses of the emitting electrons in the jet and the accretion rate onto the black hole could be at the base of the so-called ``blazar sequence" \cite{Ghisellini1998}, i.e. the  trend between the observed luminosity (progressively increasing from BL Lacs to FSRQ) and the synchrotron and high-energy SED peak frequencies (decreasing from BL Lacs to FSRQ) displayed by the blazar population \cite{Fossati1998} -- but see \cite{Giommi2012} for an alternative view.

While there is wide consensus about the  general picture, there are several fundamental questions still awaiting an answer. The list of the most pressing problems includes the nature of the mechanism powering the jet, its structure, its composition ($ep$ or $e^{\pm}$?), the role of the magnetic field. Although not conclusive, the modelling of the emission from blazars -- especially of the high-energy emission -- is used as an effective tool to start to address  several of these questions. 

The most adopted emission models (``one-zone") assume that a single region of the jet is responsible for the bulk of the observed emission.\footnote{This is strictly true for frequencies above the radio band ($\nu\gtrsim 100$ GHz). At lower frequencies it is expected that the source is opaque due to synchrotron-self absorption. Radio emission must therefore arise from larger, less compact, regions downstream of the blazar region  \cite{BlandfordKonigl1979}.} This region could be identified with the shocked portion of the jet resulting from the collision of  parts of the jet moving at different speeds (internal shocks), \cite{Spada2001}. One-zone models have the advantage to require a limited number of free parameters. The simplest version of the one-zone leptonic scenario is at the base of the the one-zone synchrotron-self Compton model, which assumes that the IC component is produced through the scattering of the synchrotron radiation produced by the same relativistic electrons. This framework is thought to be 
especially suitable to model BL Lacs, in which the possible external sources of soft target photons are negligible. The limited number of free physical parameters required by the one-zone SSC model allows us to uniquely determine them in case a well sampled SED is available \cite{BednarekProtheroe1997a, Tavecchio1998}. 

The application of the one-zone SSC model to the SED of the $\gamma$-ray emitting BL Lac (e.g., \cite{Tavecchio2010}) requires magnetic fields in the range 0.01-1 G, typical radius in the interval $10^{15}-10^{16}$ cm, Doppler factors\footnote{The relativistic Doppler factor $\delta$, determining the apparent amplification of the emission, is defined by $\delta=1/[\Gamma(1-\beta\cos\theta_{\rm v})]$, where $\Gamma$ is the jet bulk Lorentz factor, $\beta$ the jet speed and $\theta_{\rm v}$ the viewing angle.} in the interval $\delta=10-30$ and electron energies (at the SED peak) of the order of 0.1-1 TeV. If the jet is assumed to be conical, with typical aperture angle $\theta_{\rm j}\approx 5^{\rm o}$ the source radius can be translated into a distance from the central engine, $r_{\rm em}\approx 10^{16}-10^{17}$ cm, corresponding to $10^2-10^3$ gravitational radii. The results also suggest  that most of BL Lac objects are quite inefficient in emitting the radiation, since the derived cooling time of the electrons is generally much longer then the dynamical timescale and thus most of the energy stored in the relativistic particles is lost \cite{TavecchioGhisellini2016, InoueTanaka2016}. Somewhat paradoxically for source emitting conspicuous VHE radiation, a relatively low efficiency is also found to characterize the acceleration  process itself, see \cite{Baring2014}. Another important point is that the emission regions appear to be matter-dominated, the magnetic field providing a negligible contribution to the power carried by the jet (more on this later). For FSRQ the situation is more complex, due to the presence of several sources of external photons potentially involved in the IC emission \cite{GhiselliniTavecchio2009}. The ``canonical" choice is to locate the emission region within the BLR, thus exploiting for the IC emission the dense radiation field produced by the photoionized gas. Representative values of the magnetic field are larger than those of BL Lacs, in the range 1-10 G, but Doppler factors and radius are similar \cite{Ghisellini2010,GhiselliniTavecchio2015}. For FSRQ we have some control on the accretion power through the observed thermal components and it is thus possible to compare the power carried by the jet with that advected by the accreting material. The comparison show that, on average, jets carry a power larger than that associated to the infalling accretion flow, suggesting that the source of the jet power is the BH spin \cite{Ghisellini2010,Ghisellini2014}, as supported by recent GR-MHD simulations \cite{MAD}.

Although very attractive, the one-zone model is clearly quite a simplification of the actual, likely complex, structure of the emitting region(s). A minimal approach is to add one or more supplementary emission regions, such as in two-component models, e.g. \cite{BarresdeAlmeida2014, Aleksic2015} A more refined modelling within the shock-in-jet scenario includes a shock front -- at which particles are accelerated -- moving within a ``background" plasma, in which the particles injected by the shock radiatively cool and emit \cite{Kirk1998,Chen2011,Chen2015}. 
Among all possible extensions of the one-zone framework (see below for other alternatives stemming from the interpretations of the ultra-rapid variability) I would like to mention in particular the {\it structured jet} model \cite{Ghisellini2005}, which envisages a flow with two components, a faster core (the {\it spine}) surrounded by a slower sheath or layer. This kind of structure has been advanced as a possible solution for several issues related to TeV emitting BL Lacs and to unify the BL Lacs and radiogalaxy populations  \cite{Chiaberge2000, Meyer2011}.  Direct  radio VLBI imaging of jets both in low-power radiogalaxies and BL Lac objects  (e.g. \cite{Nagai2014,Giroletti2004}), often showing a ``limb brightening"  transverse structure, provides a convincing observational support to this idea, further supported by MHD simulations \cite{McKinney2006, Rossi2008}.  For this system we expect an increased IC $\gamma$-ray luminosity, based on the fact that for particles carried  by the faster (slower) region, the radiation field produced in the layer (spine) is amplified by the relative motion between the two structures \cite{Ghisellini2005, TavecchioGhisellini2008}. The spine-layer structure could be  involved in the possible production of high-energy neutrino by BL Lac jets \cite{TGG14}. The structured jet scenario can also accommodate the high-energy emission (extending at VHE) observed in the misaligned jets of radiogalaxies, such as M87 \cite {TavecchioGhisellini2008} and NGC 1275 \cite{TavecchioGhisellini2014}, although the large opacity to $\gamma$ rays caused by the intense layer radiation field could be an issue.

\section{SOME OPEN ISSUES}

\subsection{Ultra-fast variability:  sparks from magnetic reconnection?}

Vary rapid variations of the $\gamma$-ray  flux, with inferred doubling time-scales down to few minutes, have been observed in several blazars (both BL Lacs and FSRQ), most notably PKS 2155-304 \cite{Aharonian2007}, Mkn 501 \cite{Albert2007}, BL Lac \cite{Arlen2013}, PKS 1222+21 \cite{Aleksic2011}, IC 310 (\cite{Aleksic2014} although the precise classification of this source is still matter of debate) and, last but not least, 3C279 (at GeV energies by pointed observations with {\it Fermi}/LAT, \cite{Ackermann2016}). Such small timescales directly imply, via the standard causality argument, very compact emission regions, with radii not exceeding $r< c t_{\rm var} \delta\approx 10^{14} (t_{\rm var}/1 \,{\rm min}) \; (\delta /10$) cm, even smaller than the Schwarzschild radius of the black hole ($R_s=2GM/c^2=2\times 10^{14} M_{9}$ cm) and incompatible with the expected size of a shock, encompassing a large fraction of the jet cross sectional area 
\cite{Begelman2008,NarayanPiran2012}. For a recent compilation and discussions, see \cite{VovkBabic2015}.

These observational evidences stimulated new theoretical approaches aimed at modeling the structure of the emission region(s) and trying to explain the required rapid acceleration of the relativistic particles. The very compact regions implied by the rapid variability have been identified either with turbulent cells \cite{NarayanPiran2012,Marscher2014}, with large plasmoids resulting from efficient magnetic reconnection in a relativistic regime \cite{Giannios2009,Giannios2013,Sironi2015} or with shocks formed in the jet around stars crossing the jet itself \cite{Bednarek1997,Barkov2010}. In the first two cases (turbulence and reconnection), the compact regions could move relativistically in the jet reference frame (forming the so-called ``mini-jets"). In this case the resulting relativistic beaming of the emitted radiation in the observer frame (dictated by the combination of the beaming factor of the compact region in the jet frame and that of the jet flow with respect to the observer) can be very large, with Doppler factors easily reaching $\delta\sim50$. 

\begin{figure}
\hspace*{0.9 truecm}
\includegraphics[width=280pt]{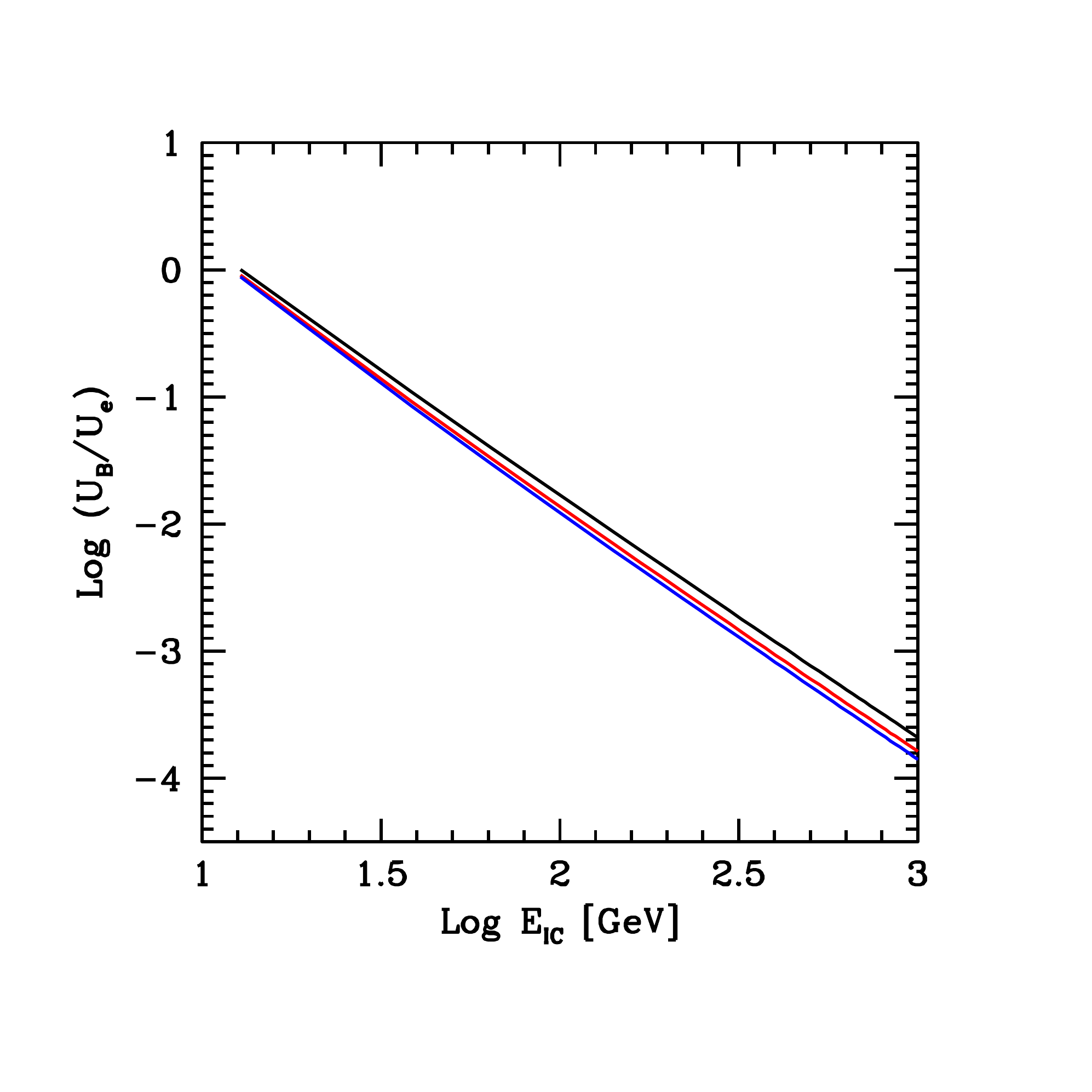}
\hspace*{-1.3 truecm}
\includegraphics[width=275pt]{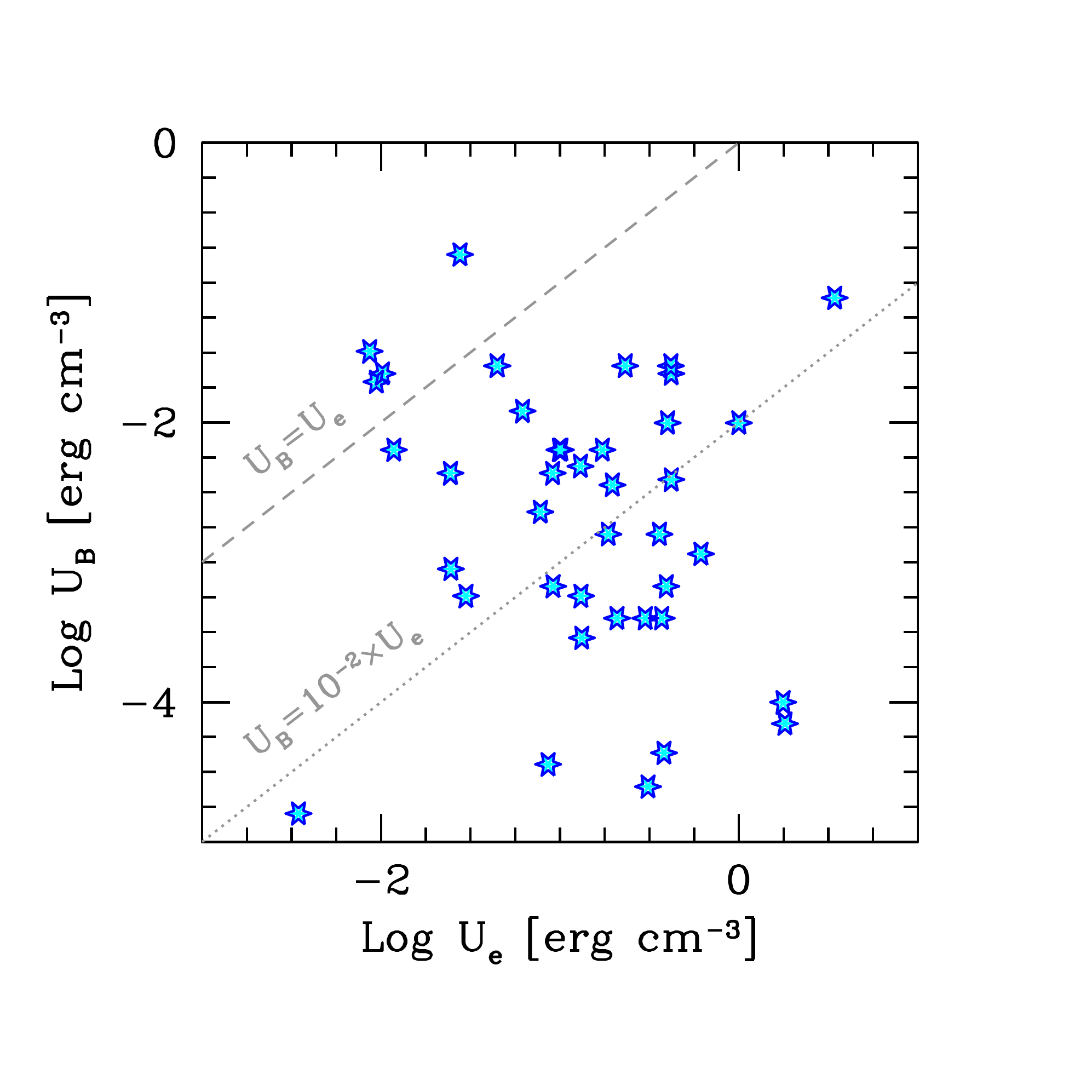}
\caption{{\it Left:} ratio between the magnetic and the relativistic electron energy density for the one-zone leptonic model as a function of the peak energy of the SSC component, for three different values of the Doppler factor, $\delta=10$ (black), 20 (red), 30 (blue) and representative values of the other spectral parameters. Typical BL Lac detected at VHE have $E_{\rm IC}\gtrsim 100$ GeV, implying $U_B/U_{\rm e}\lesssim10^{-2}$. {\it Right}: magnetic energy density, $U_B$, versus relativistic electron energy density, $U_{\rm e}$, derived for a group of $\gamma$-ray emitting BL Lac modelled with the one-zone leptonic model. Most of the sources lies in the region $U_B\ll U_{\rm e}$.  Adapted from \cite{TavecchioGhisellini2016}.}
\label{ubue}
\end{figure}

The case of IC 310 is somewhat peculiar, since its jet  is likely slightly misaligned with respect to the line of sight ($\theta _{\rm v} \simeq 10^{\rm o}$, to be compared to typical viewing angles for blazars $\theta_{\rm v}\lesssim 5^{\rm o}$) and thus the Doppler factor cannot be larger than a few, thus making the interpretation of the rapid variability (with doubling times of few minutes as observed during an active state in november 2012) even more difficult. Moreover, also the observed VHE spectrum is quite peculiar, showing an unbroken hard (photon index close to 2) power law up to 10 TeV. An exciting possibility is that the fast variations flag the development of electromagnetic cascades triggered by pairs accelerated to the required TeV energies by unscreened electric fields in gaps of the black hole magnetosphere \cite{Aleksic2014}, a mechanism already proposed to work in other low-luminosity AGN, in particular the radiogalaxy M87 \cite{NeronovAharonian2007,LevinsonRieger2011,Broderick2015}, and possibly related to the formation of the jet itself. Detailed calculations \cite{Neronov2015} show that the expected spectrum (originating from the IC scattering of the soft IR radiation from the radiatively inefficient accretion flow by the pairs in the cascade) can indeed be very hard and it is expected to cut-off around 10 TeV because of $\gamma\gamma$ absorption caused by the IR radiation field. A similar scenario involves the acceleration of narrow beams of electrons in the magnetosphere of the black hole through magnetocentrifugal effects \cite{Rieger2008,Ghisellini2009}.

\begin{figure}
\centerline{\includegraphics[width=280pt]{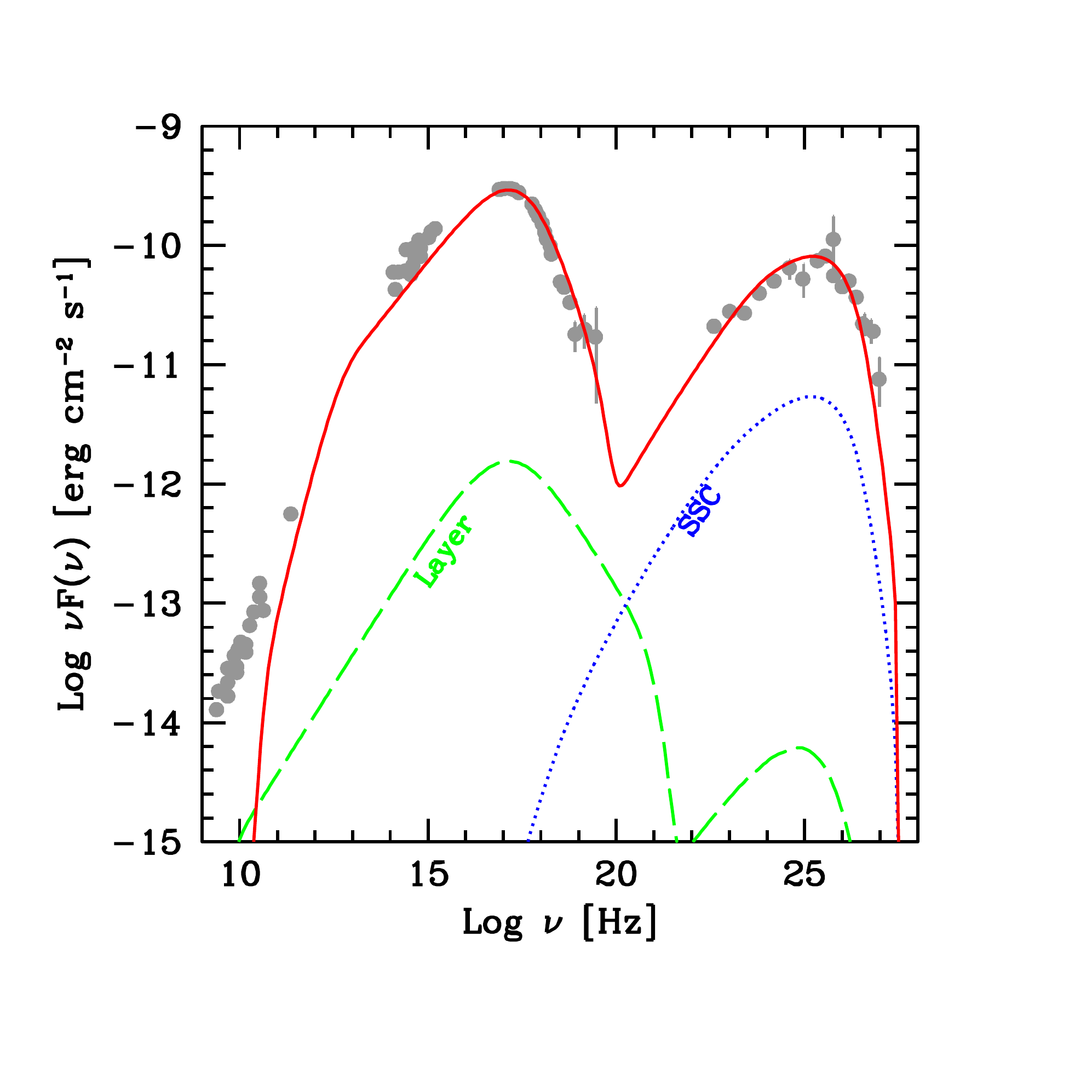}}
  \caption{SED of Mkn 421 (gray, from \cite{Abdo2011}) reproduced with the spine-layer model (red solid line). The green long-dashed line displays the assumed emission from the layer. The blue dotted line shows the SSC emission alone. As explained in the text, the large radiation energy density available for the IC emission -- provided by the layer -- increases the IC luminosity and allows one to reproduce the SED in equipartition conditions. Adapted from \cite{TavecchioGhisellini2016}.}
\label{spinelayer}
\end{figure}

Coming back to blazars, the scenario mentioned above involving the formation of plasmoids during magnetic reconnection events  (jets-in-the jet model) is being explored with quite some details and the results appear rather promising (see Sironi, these proceedings). From a fundamental  perspective, magnetic reconnection-powered emission from jets is favored with respect to that associated to shocks since -- as expected at the typical distance of the ``blazar zone" -- the flow is thought to be largely dominated by the Poynting (i.e. magnetic) flux \cite{Komissarov2007}. In these circumstances diffusive particle acceleration by shocks is expected to be quite inefficient \cite{Sironi2015,Sironi2015b}, while the large energy stored in the magnetic field can be efficiently extracted and channeled to particles. Recent dedicated particle-in-cell plasma simulations \cite{SironiSpitkovsky2014} demonstrate  that the combined action of the acceleration by unscreened electric fields in the current sheets and the subsequent Fermi I-like acceleration caused by the bouncing of the particles off the magnetic islands, provides an efficient and rapid way to accelerate electrons (and ions). Simulations further robustly show that the post--reconnection regions -- in which most of the radiation we receive is produced -- should be characterized by a substantial equipartition between the relativistic particles and the magnetic field.

Some problems for the magnetic reconnection scenario arises when the general predictions sketched above are compared to the physical state of the jets (or, more precisely, of the emission regions of the jets) that we infer by modelling the SED. First of all it appears that, contrary to the assumptions of the model, the jets of VHE BL Lacs (which are the majority of the sources in which ultra-rapid variability events have been observed) are rather weakly magnetized, with the Poynting flux providing a negligible contribution to the total jet power (incidentally, this could be a major problem also for the general scenario for jet acceleration that foresees a substantial equality of magnetic and kinetic luminosities, their ratio decreasing only logarithmically with distance after the acceleration phase). Similarly, the regions probed through the emission are largely far from equipartition conditions, with derived magnetic over electronic energy density ratio of the order of $U_B/U_{\rm e} \simeq 0.1-0.01$ \cite{TavecchioGhisellini2016}. These general conclusions are quite robust, at least in the framework of one-zone models, since, as already stressed, in this case all parameters are fixed by well sampled SED. Some works in the past already reached these conclusion on single sources, e.g. \cite{Abdo2011, Acciari2011}. In \cite{TavecchioGhisellini2016} we have extended  the study to a large group of $\gamma$-ray emitting BL Lacs. The main results are shown in Fig. 1. In the left panel we report $U_B/U_{\rm e}$ versus the energy of the high-energy SSC peak, assuming representative values for the other spectral parameters. Since most of the VHE BL Lac have the peak at energies $E_{\rm IC}\gtrsim 100$ GeV, the plot show that $U_B/U_{\rm e}\lesssim0.01$. I remark that the lines in the plot have been derived analitically. The right panel in Fig. 1show instead the derived $U_B/U_{\rm e}$ applying the one-zone SSC model to a group of BL Lac with information on the high-energy peak. As expected, most of the sources lies in the region with $U_B \ll U_{\rm e}$.

The issue related to the low magnetization of the BL Lac jets shows up also when synthetic SED are derived in the framework of the reconnection model \cite{Petro2016}. Due to the large magnetic field, the derived SED systematically display a luminous synchrotron component, much more powerful (by almost one order of magnitude) than the SSC one. This has to be compared to the observed SED which generally show a substantial equality of the two peak luminosities. The reason for the discrepancy is clearly related to the large magnetic energy density predicted as a leftover of the reconnection process, which implies a large synchrotron emissivity. 

A possible way out to these problems, advanced in \cite{TavecchioGhisellini2016}, is to invoke the spine-layer structure for the jet. The idea at the base of this scenario can be simply grasped. In fact, as already remarked, with the spine-layer structure the luminosity of the IC bump produced by the spine can be increased because of the supplementary soft target radiation energy density provided by the boosted layer emission. Therefore, the required SSC luminosity can be produced by a smaller number of relativistic  electrons, leading to reduce $U_{\rm e}$. At the same time, to maintain the same synchrotron luminosity with a reduced number of electrons one has to increase the magnetic field, that is increase $U_{B}$. The ratio between  $U_{B}$ and $U_{\rm e}$ therefore increases. The equipartition can be reached assuming a reasonable level for the layer emission, see in Fig. 2 the case of Mkn 421. Besides solving the problem of the equipartition, the larger magnetic energy density and smaller particle density  allowed by the spine-layer set-up, lead to increase the Poynting flux. In the benchmark case of Mkn 421 reported in Fig. 2, it comes out that the system reaches the condition $L_B\approx L_{\rm kin}$, thus mitigating the problem represented by the small jet magnetization.

\subsection{Flat spectrum radio quasars at VHE}

The great majority of blazars detected at TeV energies are BL Lacs (more precisely HBL -- highly-peaked BL Lacs -- defined as having the maximum of the synchrotron component in the UV-X-ray band, \cite{PadovaniGiommi1995}).
For  a long time, FSRQ have not been considered good target for VHE observatories. There are several reasons supporting such theoretical prejudice: {\it i}) FSRQ are characterized by soft MeV-GeV $\gamma$-ray spectra (their SED peaks are generally located at $E<100$ MeV); {\it ii}) the continuum, if produced through IC scattering, is expected to further soften above few GeV due to effects related to the Klein-Nishina cross section, \cite{TavecchioGhisellini2008b}; {\it iii}) another important factor to consider is the anticipated huge ``internal" absorption of $\gamma$-ray photons with energies exceeding few GeV interacting through the $\gamma\gamma\to e^{\pm}$ reaction with the various soft radiation fields in the FSRQ nucleus -- most notably that associated to the BLR \cite{Donea2003, LiuBai2008}. In fact, $\gamma$ rays produced well inside the BLR are characterized by optical depth $\tau_{\gamma\gamma}\gg1$ ( i.e. absorption probabilities close to 1) for energies above 20 GeV -- the energy threshold for the interaction with the abundant photons of the hydrogen Lyman $\alpha$ broad emission line. 

Despite the pessimistic expectations, photons at several tens of GeV (source rest frame) from FSRQ are occasionally detected by {\it Fermi}-LAT, usually during high-states or flare \cite{Pacciani2014,Ackermann2016}. Generally, during these events the $\gamma$-ray spectra become quite hard, with photon index close to 2 (i.e. ``flat" in the SED). These episodes are also commonly (but not always, e.g. \cite{Tavecchio2011}) accompanied by a shift of the entire SED toward higher frequencies, hence the name ``blue flat spectrum radio quasars" \cite{Ghisellini2012,Ghisellini2013} sometimes used to describe FSRQ with these peculiar SED. 

\begin{figure}
\centerline{\includegraphics[width=270pt,angle=90]{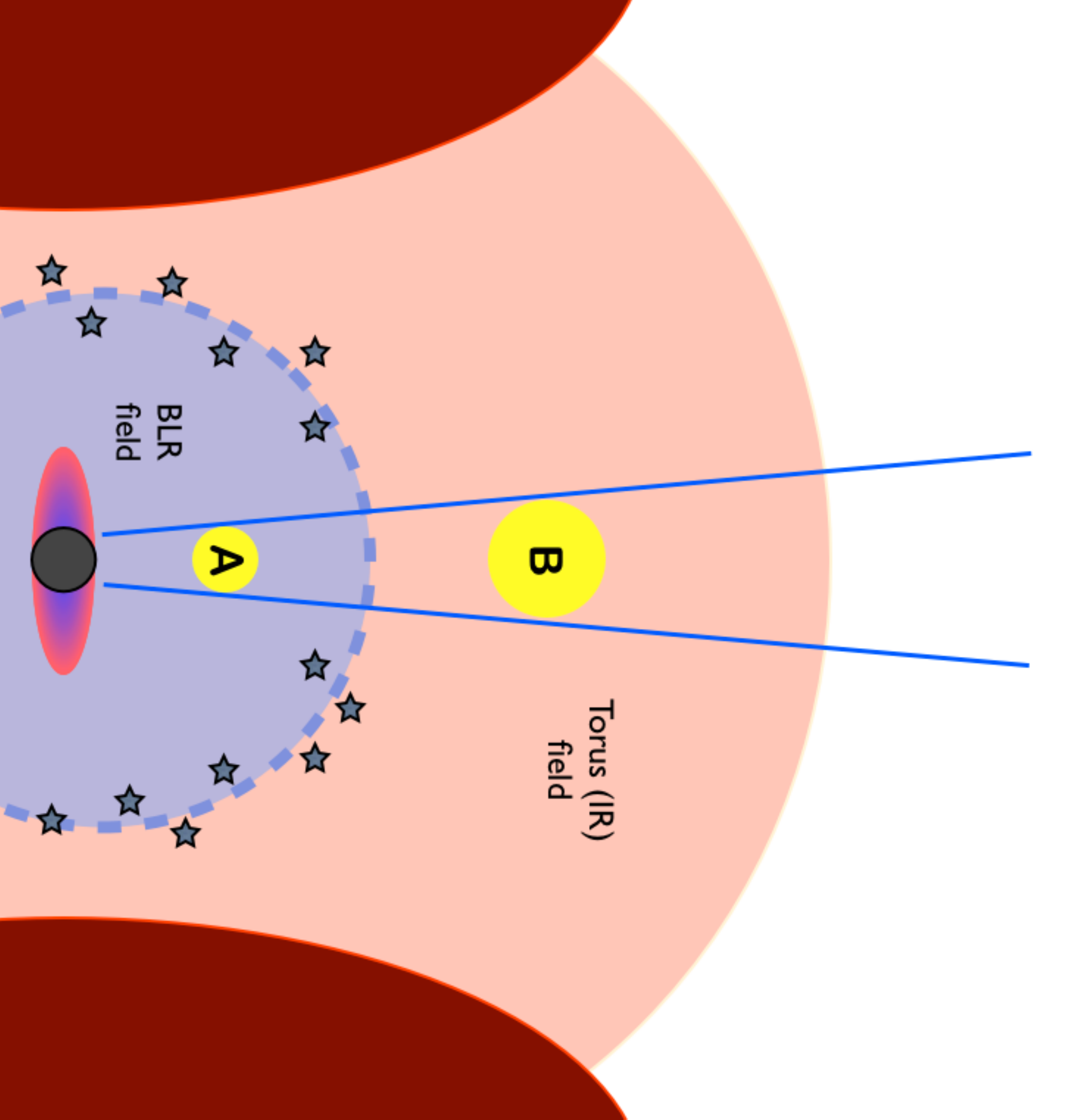}}
  \caption{Cartoon of the structure of a FSRQ (not to scale). Standard models assume that the bulk of the emission occurs within the BLR, at distances $\lesssim 0.1$ pc (case A). To avoid huge absorption of $\gamma$ rays the emission region must be located outside the BLR radiation field (case B), where the external radiation field is dominated by the IR thermal emission of the pc-scale dusty torus (pink).}
\label{cartoon}
\end{figure}

To avoid the huge opacity to high-energy $\gamma$ rays, the emission during these events should occur outside the BLR. These more or less temporary states could thus be interpreted as (rare?) phases during which the emission region moves from within the BLR (where it is thought to permanently reside in classical ``red" FSRQ) to the outer regions, where the external radiation field is 
dominated by the IR radiation field of the dusty torus \cite{Pacciani2012, Ghisellini2013} (see Fig. \ref{cartoon}). 
In \cite{GhiselliniTavecchio2008}, along the lines of \cite{Georganopoulos2001}, we remarked that the FSRQ most likely susceptible to change their SED from ``red" to ``blue" are those characterized by a relatively small size of the BLR, so that the blazar emission zone can easily go beyond it even in case of relatively small displacements. Since, as we discussed above, the BLR radius is set by the disk luminosity (as $r_{\rm BLR}\propto L_{\rm d}^{1/2}$), the sources mostly prone to such transitions are expected to be FSRQ of low power. 

The effect of the displacement of the emission region out of the BLR radiation field is twofold. On one hand the reduced optical depth permits the propagation of photons above 20-30 GeV, as observed during flares. On the other hand, a reduction of the external radiation energy density -- which determines the cooling rate of the electrons, dominated by the IC process off the external photons, $\dot{\gamma}_{\rm cool, IC}\propto U_{\rm rad}^{-1}$ -- implies a reduced radiative cooling rate of the electrons and -- in the view in which the maximum electron energy is fixed by the balance between losses and gains -- a larger average energy of the accelerated electrons. The increased Lorentz factors of the electrons can then explain the shift of the SED peak at larger energies \cite{Ghisellini2013}. Of course, in these circumstances the detection of FSRQ also at VHE becomes a possibility.

\begin{figure}
\centerline{\includegraphics[width=290pt]{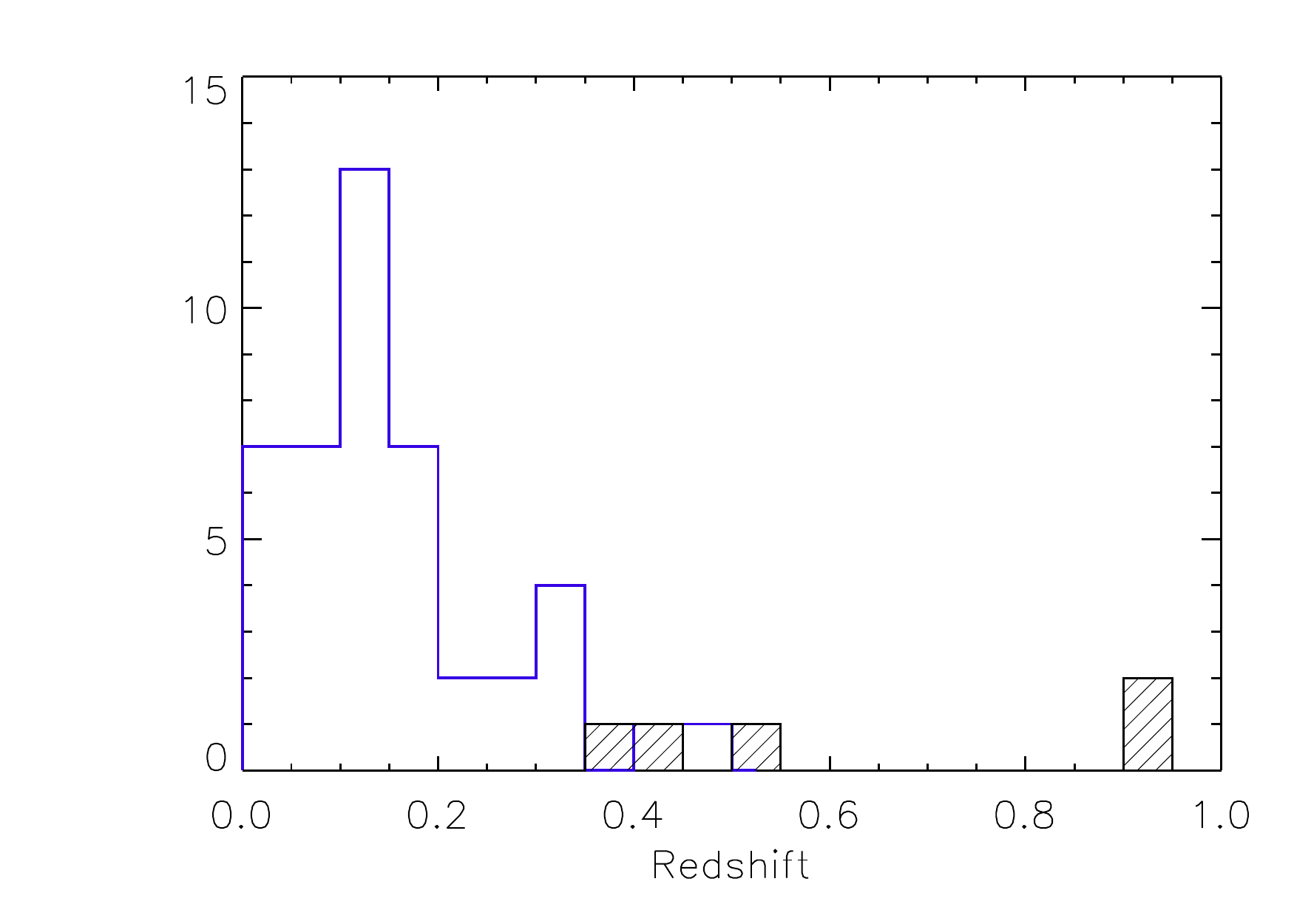}}
  \caption{Redshift distribution of BL Lacs (blue) and FSRQ (black) detected at VHE. Sources without a firm redshift determination are not included.}
\end{figure}

Until now, five FSRQ (3C 279, PKS 1510-089, 4C+21.35, PKS 1441+25, S3 0218+35)\footnote{I do not include in this list the blazar S4 0954+65, often considered a FSRQ, since it is likely a BL Lac of unknown redshift \cite{Landoni2015}.} have been detected at VHE, out of a total of 62 blazars\footnote{Data from {\tt http://tevcat.uchicago.edu}}. FSRQ and BL Lacs display a rather different cosmological evolution (for studies in the $\gamma$-ray band see \cite{Ajello2012,Ajello2014}) and this naturally reflects in the redshift distribution of the VHE detected sources (Fig. 4), with FSRQ showing redshifts significantly larger than those of BL Lacs:  in particular, PKS1441+25 and S30218+35 are located at a redshift close to 1. The large distance exacerbates the effects of the absorption through the interaction with the extragalactic background light (EBL) through the pair production reaction $\gamma + \gamma_{\rm EBL}\to e^{\pm}$. A consequence is that the recorded VHE spectra of FSRQ  are quite soft and therefore instruments characterized by a low energy threshold are favored in their detection.

As discussed above, the detection of photons with energies in the 100 GeV-1 TeV range allows us to locate the emission region beyond the BLR edge. A conservative possibility \cite{BoettcherEls2016} is to assume that the emission occurs just beyond the BLR, where the radiation field is already dominated by the IR radiation field from the dusty torus \cite{Blazejowski2000}. The lower energy of the IR photons ($\epsilon<1$ eV) has two effects: i) the KN effects on the produced EC radiation starts at larger energies \cite{Pacciani2014}; ii) the absorption threshold moves at larger energies, allowing the unattenuated propagation of $\gamma$-ray photons of hundreds of GeV.  For typical temperatures of the dust ($T\approx 10^3$ K), one expects huge absorption of the emitted $\gamma$-ray photons with $E\gtrsim 1$ TeV. This result can be derived though a simple back-on-the-envelope calculation. Close to threshold, the typical photon energy of the (black body-like) IR radiation field interacting with the $\gamma$-ray photons  is $\langle\epsilon \rangle \approx 2.82 \, kT\simeq 0.25 \, T_3$ eV, implying that absorption becomes relevant for photons of energy $E=m_{\rm e}^2c^2/\langle\epsilon \rangle = 1 \, T_3^{-1}$ TeV. The corresponding optical depth is expected to be very large. This can be well approximated as $\tau_{\gamma\gamma}=\frac{\sigma_{\rm T}}{5}\, n_{\rm ph} r_{\rm IR}$ \cite{DondiGhisellini1995}, where the $n_{\rm ph}$ is the number density of the target soft photons and $r_{\rm IR}$ is the torus size. Using a scaling law for the torus radius as a function of the disk luminosity $r_{\rm IR}=1 L_{\rm d, 45}^{1/2}$ pc \cite{GhiselliniTavecchio2009} one easily derives $\tau_{\gamma\gamma} \simeq 300 \, \xi \, L_{\rm d, 45}^{1/2} T_3^{-1}$, where $\xi\approx 0.5$ is the fraction of the disk luminosity reprocessed into IR radiation. With such a large optical depth one naturally expects an abrupt cut-off of the high-energy spectrum at energies close to 1 TeV (source frame). 

\begin{figure}
\centerline{\includegraphics[width=270pt]{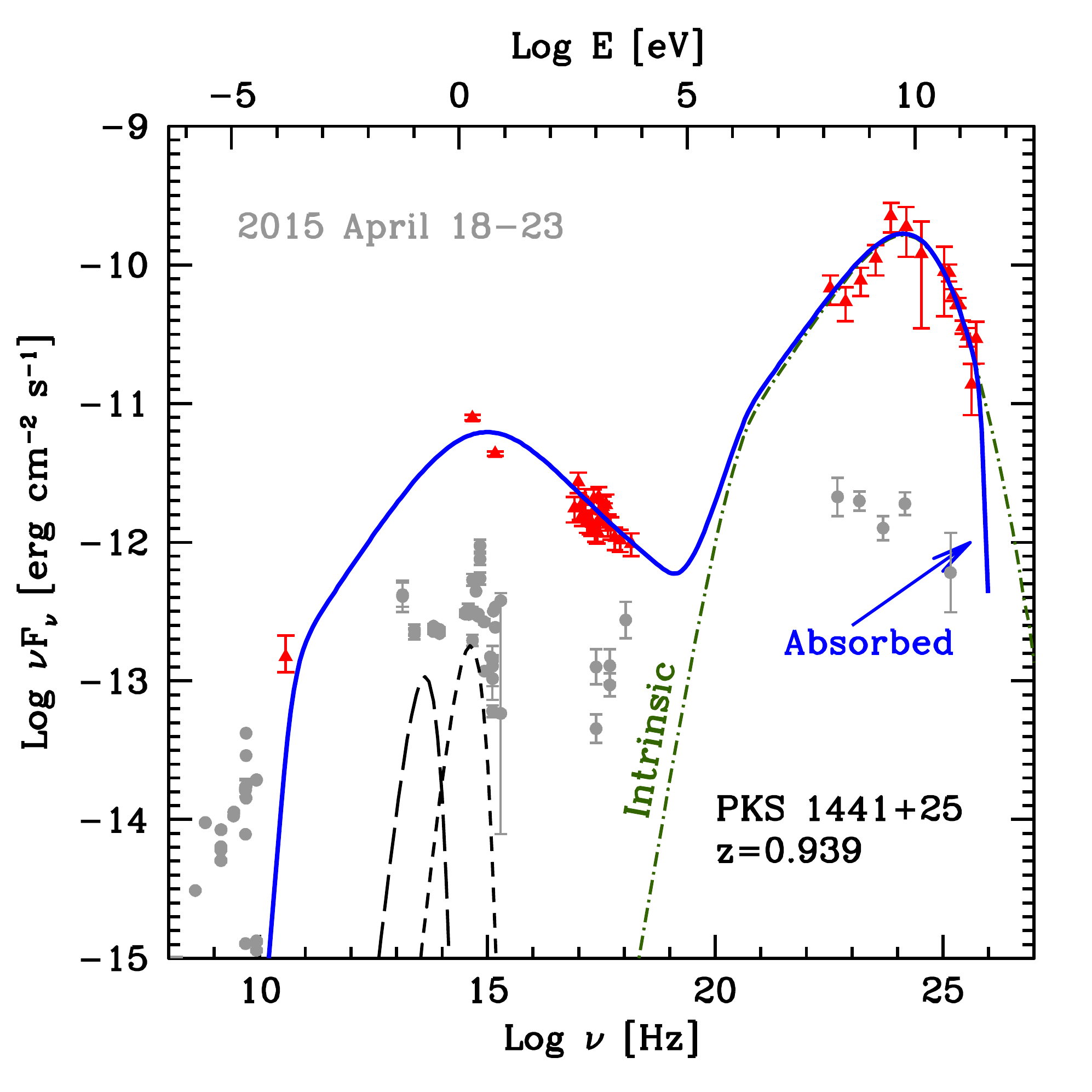}}
  \caption{SED of the FSRQ PKS 1441+25, detected at VHE by MAGIC and VERITAS. Red symbols (data taken from \cite{Ahnen2015}) refer to an active phase on 2015 April 18-23. Gray symbols show historical data for the quiescent state. We report the results of a leptonic one-zone model, assuming that the emission occurs beyond the BLR edge (blue solid line). The green point-dashed line shows the inverse Compton component (produced through the scattering of the IR photons from the dusty torus), not including $\gamma\gamma$ absorption. The strong cut-off at around 1 TeV caused by absorption is clearly visible comparing the blue and the green lines. See \cite{Ahnen2015} for  details.}
\label{1441}
\end{figure}

Unfortunately current detections do not reach such high energies. The most interesting case is that of the recently detected PKS1441+25 ($z=0.940$). The observed power-law spectrum \cite{Ahnen2015} extends up to $E_{\rm max}\approx 300$ GeV, which, in the rest frame of the source corresponds to $E^{\prime}_{\rm max}= E_{\rm max} (1+z)\approx 580$ GeV,  just at the edge expected cut-off. This is clearly visible in Fig. \ref{1441}, reporting (red symbols) the nearly simultaneous multifrequency data recorded in the period 18-23 April 2015, during the highest $\gamma$-rat state. The blue line shows the result of the one-zone model assuming that the emission region lies just beyond the BLR radius. From the luminosity of the observed emission lines one can infer that the disk luminosity is $L_{\rm d}\simeq 2\times 10^{45}$ erg s$^{-1}$ and from it, using the scaling law of \cite{GhiselliniTavecchio2009} one can infer the radius of the BLR to be $r_{\rm BLR}=1.4\times 10^{17}$ cm. The emission region is located at $r_{\bf em}=5\times 10^{17}$ cm. At this distance the (de-beamed) radiation field from the BLR is negligible and the IC emission is dominated by the scattering of the  IR radiation of the dusty torus. Clearly, the evidence for the expected absorption cut-off would be quite important, allowing to firmly locate the emission region within the IR torus. Of course, even the absence of the cut-off would be a quite precious information, demonstrating that, at least occasionally, the emission can occurs quite far out along the jet, as argued by several groups based on the observed behavior of the optical polarization, the ejection of new VLBI components apparently simultaneously with the flares or on the modelling of the SED \cite{Marscher2008,Marscher2010,Aleksic2014b,Tavecchio2013,Pacciani2014}.

The scenario assuming that $\gamma$-ray production occurs beyond the BLR naturally foresees relatively long variability timescales ($\approx$ days). In fact, in the standard cone geometry for the jet, at a distance $r_{\rm em}$ the expected jet radius is $r_{\rm j}=\theta _{\rm j} r_{\rm em}$, where $\theta _{\rm j}\approx 0.1$ rad is the jet semi-aperture angle. Using standard relations for the BLR radius one gets $r_{\rm em}>r_{\rm BLR}\approx 0.1$ pc, implying $r_{\rm j}\gtrsim 3\times 10^{16}$ cm, and thus variability timescales of the order of $\Delta t_{\rm var}> r_{\rm j}/c\delta\simeq 1 (\delta/10)^{-1}$ days. This line of reasoning was used to argue that the evidence of $\approx$ hour timescale variability implies that the emission occurs at small distances, well within the BLR \cite{Tavecchio2010b}. Indeed, the variability timescale inferred for most of the FSRQ detected at VHE, as PKS 1441+25 discussed above, show relatively smooth and slow variations in $\gamma$-rays. However there are some cases in which the condition is evidently violated. The first case was that of 4C+21.35, for which MAGIC detected a flare with a raising time of just 10 minutes \cite{Aleksic2011}, readily implying a quite compact emission region $r_{\rm em}\approx 10^{14} (\delta/10)$ cm located beyond the BLR \cite{Tavecchio2011}. Similar is the recent case of the flare of 3C279 detected by LAT \cite{Ackermann2016}, which reported variability with $\Delta t_{\rm var}\approx 5$ minutes accompanied by the detection of photons with energies exceeding 50 GeV (likely flagging emission beyond the BLR). 

The possible scenarios trying to reconcile minute-scale variability (hence requiring small emission regions, with $r_{\rm em}\approx 10^{14}$ cm) with the absence of absorption (thus locating the emission region at large distances, outside the BLR) follow the lines of those already discussed above for the ultra-fast variability. For FSRQ, the physical conditions inferred for the emission regions are in better agreement with the magnetic reconnection scenario than those of BL Lac jets. In fact, equipartition between magnetic fields and relativistic particles (predicted for the magnetic reconnection model) generally holds for FSRQ \cite{Ghisellini2010}. However, a still acute problem could remain, since  the jets do not appear to be  magnetically dominated at the distances assumed for the emission \cite{Ghisellini2014,Nalewajko2014}, a {\it conditio sine qua non} for the model to work (magnetic reconnection models generally assume that the magnetic luminosity of the jet is $>$10 times larger than the kinetic one). However, this conclusion could be revised in case the jet composition includes at least 10-20 pairs every proton \cite{Pjanka2016}. 

Further advances in the understanding of VHE emission from FSRQ requires more data and an enlarged sample of sources. For FSRQ, even more than BL Lac objects, the most effective way to understand the underlying physics is to put the VHE data in a broader context, including radio (especially VLBI monitoring, providing precious spatial information on the activity), optical (including polarization measurements), X-rays and GeV data (where most of the action takes place). Surely this will be a fruitful field in the upcoming CTA era.

\section{OUTLOOK: TOWARD THE CTA ERA}

The field of blazar $\gamma$-ray astrophysics is living a quite exciting time. {\it AGILE} and {\it Fermi}-LAT provide an unprecedented dynamical view of these sources over several timescales. Cherenkov arrays on ground are revealing an unexpected variety of phenomena, such as the ultra-rapid flares and the very-high energy emission of FSRQ. The current observational evidence is probably just the top of the iceberg. 

In particular, the current sparse observations of ultra-rapid variability events in the VHE band leaves several questions unanswered. In particular the duty cycle (i.e. the frequency) of these events is a key information, potentially able to test and to constrain some of  the proposed scenarios. Of course, a proper evaluation of the duty cycle would require a dense and frequent monitoring of the most promising targets. The events detected so far happens during high-state, flaring states of the sources. It would be quite interesting to know whether this kind of phenomena can also occur in states of low/quiescent activity. Particularly relevant for our understanding would be the detection of such events in other bands, such as optical or X-ray. So far, observations have not revealed a phenomenology as that observed in the high-energy band. To check whether ultra-rapid variability appears also in these bands, strictly simultaneous observations in different bands are required.

The detection of very-high energy photons from FSRQ is a quite informative result already changing our view of these sources. We know that, at least occasionally, the emission region of jets in FSRQ can leave the BLR. We do not have a clear explanation of the mechanisms triggering these states and how often they occur. The extension of the detection within the TeV band could provide other important information, especially if a cut-off induced by absorption on the IR radiation field is revealed. In principle such observations would even provide us the opportunity to probe the IR emitting regions of the AGN.

Although in this paper I focused on the two issues discussed above, it is important to mention two other topics potentially very important related to the high-energy emission from blazars. The first topic is that involving the so-called {\it extreme} BL Lacs, characterized by several peculiar aspects, such as an extreme energy of the synchrotron peak maximum (in some cases exceeding 10 keV), an exceptionally hard GeV-TeV continuum, locating the IC peak above several TeV (to be compared with a typical IC peak energy of 100 GeV) a limited variability at high-energy. An interesting scenario interprets these peculiarities invoking the existence of a beam of ultra-high energy hadrons escaping from the jet and producing the $\gamma$-ray emission on the way toward the Earth through Bethe-Heitler and photo-meson cooling \cite{Essey2011}. A robust and testable prediction of this model is the presence of a detectable hard $\gamma$-ray tail extending above 10 TeV \cite{Murase2012} even for sources located at redshift close to 1 \cite{Aharonian2013}. In the standard models the EBL absorption would prevent the detection of photons with these extreme energies. A detection of even few photons with $E>10$ TeV by CTA would therefore be a spectacular confirmation of the hadronic beam model.

Finally I would like to mention the potential role of blazars as cosmological tracers of the evolution of the most supermassive black holes. As stressed by \cite{Ghisellini2010SMBH} the most powerful FSRQ are indeed hosted in systems with a mass of the black hole exceeding $10^9$ solar masses. The detection of these FSRQ at high redshift (beyond 2-3) is thus a powerful tool to make a census and follow the evolution of massive black holes \cite{Ghisellini2013SMBH}. Powerful FSRQ display an IC peak with a maximum in the soft gamma-ray band, around few MeV. Indeed they usually escape detection by {\it Fermi}-LAT (characterized by a low energy limit of 100 MeV) but appear in hard X-ray surveys such as that performed by BAT onboard {\it Swift}. A powerful detector of MeV blazars such as the proposed {\it e-ASTROGAM} woulf be the ideal instrument for this kind of studies.


\section{ACKNOWLEDGMENTS}

I thank the organizers for the invitation. It is a pleasure to thank G. Ghisellini and G. Bonnoli for years of fruitful collaboration. I am grateful to L. Sironi for stimulating discussions. This work has been partly founded by a PRIN-INAF 2014 grant.



\end{document}